\documentclass[aps,prl,reprint,groupedaddress,superscriptaddress]{revtex4-1}
\usepackage{amsmath}
\usepackage{amssymb}
\usepackage{amsfonts}
\usepackage[dvipdfmx]{graphicx,color}
\begin{document}

\title{Impact-induced energy transfer and dissipation in granular clusters under microgravity conditions}

\author{Hiroaki Katsuragi}
\affiliation{Department of Earth and Environmental Sciences, Nagoya University, Furocho, Chikusa, Nagoya 464-8601, Japan}

\author{J\"urgen Blum}
\affiliation{Institut f\"ur Geophysik und extraterrestrische Physik, Technische Universit\"at zu Braunschweig, Mendelssohnstr. 3, D-38106 Braunschweig, Germany}

\date{\today}

\begin{abstract}
The impact-induced energy transfer and dissipation in granular targets without any confining walls are studied by microgravity experiments. A solid projectile impacts into a granular target at low impact speed ($0.045 \leq v_p \leq 1.6$~m~s$^{-1}$) in a laboratory drop tower. Granular clusters consisting of soft or hard particles are used as targets. Porous dust agglomerates and glass beads are used for soft and hard particles, respectively. The expansion of the granular target cluster is recorded by a high-speed camera. Using the experimental data, we find that (i)~a simple energy scaling can explain the energy transfer in both, soft- and hard-particles granular targets, (ii)~the kinetic impact energy is isotropically transferred to the target from the impact point, and (iii)~the transferred kinetic energy is $2$~-~$7$\% 
of the projectile's initial kinetic energy. The dissipative-diffusion model of energy transfer can quantitatively explain these behaviors.
\end{abstract}

%\pacs{}
%\keywords{}

\maketitle

In spite of its simplicity, granular matter offers various intriguing phenomena~\cite{Andreotti:2013}. Impact response is one of the most fundamental problems in granular mechanics~\cite{RuizSuarez:2013,Katsuragi:2016,vanderMeer:2017}. For instance, impact drag force~\cite{Katsuragi:2007,Katsuragi:2013,Clark:2014,Bester:2017}, cratering~\cite{Walsh:2003,Wada:2006}, and splashing~\cite{Crassous:2007,Valance:2009,Deboeuf:2009,Tanabe:2017} have been extensively studied from the viewpoint of soft matter physics. In these studies, macroscopic particles~($>0.1$~mm in diameter) have been mainly used to neglect the effects of interstitial air drag and cohesion among particles. Cohesion can only cause the weak clustering in free-falling granular flow~\cite{Royer:2009}. For fine particles ($\sim 1$~$\mu$m), the cohesive force is strong enough to keep macroscopically stable porous structures without any support by a container. Since such porous dust agglomerates are considered the building blocks in the planet formation process, their mechanical properties and collisional outcomes have been studied in the context of planetary science~\cite{Blum:2018}. Motivated by both, granular physics and planetary science, impacts of solid projectiles into porous-dust agglomerate have been experimentally investigated~\cite{Guttler:2009,Katsuragi:2017}. In addition, the collisional outcome of {\it hierarchical granular clusters} ($\sim 1$~cm), which consist of macroscopic particles ($\sim 1$~mm) that are porous-dust agglomerates made of fine particles ($\sim 1$~$\mu$m), has also been studied recently~\cite{Whizin:2017}. The $1$-mm-sized porous-dust agglomerates are much softer than usual hard particles, such as glass beads. This difference could affect the impact dynamics. However, no systematic comparison between soft (porous-dust-agglomerates) granular clusters and hard (glass-beads) granular clusters have been carried out in terms of their mechanical characterization and impact response, although it is mandatory for an in-depth understanding of both, granular physics and planet formation.

To reveal the general feature of granular impacts, the impact response of hierarchical (soft) granular clusters have to be examined and compared to that of dense (hard) granular clusters. One has to be careful when studying the physics of hierarchical granular clusters. First, the interstitial air effect must be removed, since the smallest units of the hierarchical granular clusters are micrometer in size and, thus, too small to neglect air drag. Besides, the effect of gravity should be reduced to observe the intrinsic granular nature without gravitational-loading and confining-wall effects for considering astrophysical application~\cite{Murdoch:2017,Planes:2017}. 
For this purpose, we performed a solid-projectile-impact experiment in a vacuum laboratory drop tower, using both, soft- and hard-particle clusters. By analyzing the experimental results, general features of granular impact, free of interstitial-air and gravitational effects are presented and discussed in this paper. Particularly, the impact-induced energy transfer and dissipation are analyzed on the basis of image analysis. To the best of our knowledge, this is the first report on the impact dynamics of free-falling hierarchical granular clusters.

The basic concept and principle of the laboratory drop tower we used were presented by~\cite{Blum:2014}. The system used in this study is identical to that of~\cite{Whizin:2017}. Here, we briefly summarize the experimental setup. Porous dust agglomerates were prepared by sieving a powder consisting of polydisperse ($0.1$~-~$10$~$\mu$m) SiO$_2$ monomers of irregular shape~(Sigma-Aldrich). During the sieving process, the monomers readily stick to each other due to cohesive forces and form macroscopic porous particles. In this experiment, we used porous dust agglomerates of diameters ranging from $1.0$ to $1.6$~mm. For comparison, spherical glass beads~($1$~mm in diameter) were also used as granular particles. Dust agglomerates or glass beads were poured into a cup of diameter $25$~mm where they formed a soft or hard granular cluster target. The mass of the granular target was $1.0 \pm 0.1$~g for soft (dust-agglomerate) clusters and $3.0 \pm 0.1$~g for hard (glass-beads) clusters. The granular clusters in the cup and the solid projectile are installed in the top of the drop tower. The centers of mass of projectile and target were aligned in vertical direction and were released by removing the cup fast enough to attain free fall without any initial velocity. By controlling the release times of projectile and target, collisions with impact speeds $v_p$ in the range of $0.045 \leq v_p \leq 1.6$~m~s$^{-1}$ can be achieved during the free fall. Since the impact occurs during free fall, the effect of Earth's gravity, the weight-force $mg$, becomes irrelevant in this free-fall system with acceleration $g$, because it is perfectly compensated by a fictitious force $-mg$. Here, $m$ and $g$ are the mass of the object and Earth's gravitational acceleration. A high-speed camera was simultaneously released at the instance of target release to capture the impact dynamics with $3,000$ frames per second and $0.21$~mm per pixel spatial resolution. The residual pressure in the drop tower was kept at $6$~Pa. 

\begin{figure}
\begin{center}
\resizebox{0.45\textwidth}{!}{\includegraphics{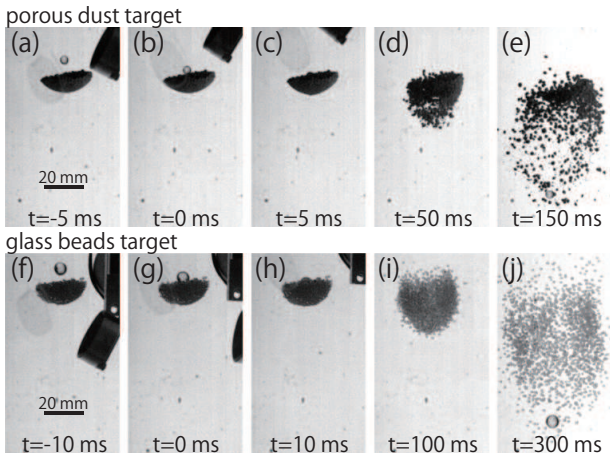}}
\end{center}
\caption{The top row (a-e) shows the impact of a $D_p=4$~mm glass projectile into a porous-dust-agglomerate cluster at $v_p=0.82$~m~s$^{-1}$ impact speed. The bottom row (f-j) shows the impact of a $D_p=6$~mm glass projectile into a dense-glass-bead cluster at $v_p=0.38$~m~s$^{-1}$ impact speed. In both cases, impact-induced expansion of the target cluster followed by fragmentation can be observed. The dark parts at the top-right corners in panels (a-c, f-h) show the target-release system.} 
\label{fig:raw}
\end{figure}

We employed several types of spherical solid projectiles: glass beads ($D_p=4$, $6$, or $10$~mm, and $\rho_p=2.5 \times 10^3$~kg~m$^{-3}$), lead beads ($D_p=4.5$~mm and $\rho_p=11 \times 10^3$~kg~m$^{-3}$), and plastic beads ($D_p=6$~mm and $\rho_p=0.99 \times 10^3$~kg~m$^{-3}$), where $D_p$ and $\rho_p$ are the diameter and density of the projectile, respectively. In total, 64 impacts (36 for porous-dust agglomerate targets and 28 for glass-beads targets) were performed. Example images taken by the high-speed camera are shown in Fig.~\ref{fig:raw}, in which $t$ is the elapsed time since the impact moment.

The top and bottom rows in Fig.~\ref{fig:raw} display porous-dust-agglomerate and dense-glass-bead cluster cases, respectively. A glass projectile is used in both cases. In these impacts, the kinetic energy is sufficiently large to break the target cluster. Clear fragmentation can be observed in the late stage~(Fig.~\ref{fig:raw}(e,j)). For very low impact velocities, however, the drop tower is too short (the available time is about $0.5$ s) to observe the late-stage fragmentation dynamics. Therefore, in this study, we will focus on the kinematics and early-stage expansion dynamics of the target clusters.

\begin{figure}
\begin{center}
\resizebox{0.45\textwidth}{!}{\includegraphics{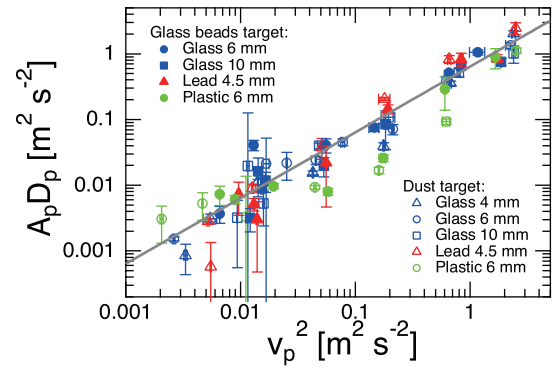}}
\end{center}
\caption{Kinematic scaling relation between the characteristic acceleration $A_p$, the impact velocity $v_p$, and the projectile diameter $D_p$. The gray line indicates the scaling $A_p D_p = C v_p^2$ with a fitting parameter $C=0.64$.}
\label{fig:ApDp-vp}
\end{figure}

\begin{figure}
\begin{center}
\resizebox{0.45\textwidth}{!}{\includegraphics{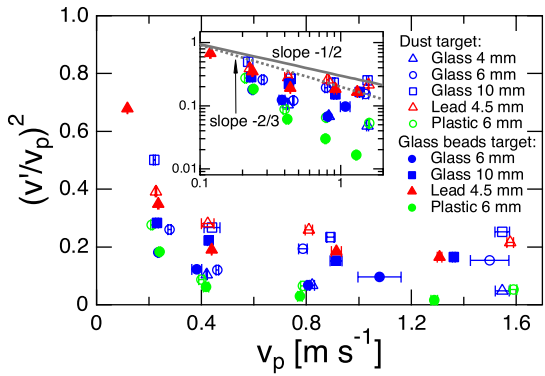}}
\end{center}
\caption{The normalized projectile energy after the break-through of the target, $(v'/v_p)^2$, as a function of impact velocity $v_p$. Inset shows the same data in a log-log plot. The two lines correspond to the viscoelastic (slope -2/3) and plastic (slope -1/2) behaviors.}
\label{fig:after-vel}
\end{figure}

First, we analyze the projectile kinematics. By image analysis, the relative impact velocity, $v_p$, and the deceleration due to the impact, $A_p$, were measured~(see appendix for the method of image analysis). Similarly to~\cite{Katsuragi:2017}, a simple kinematic coupling between $v_p$, $A_p$, and $D_p$ is derived as shown in Fig.~\ref{fig:ApDp-vp}. From the data behavior, one can confirm the relation
\begin{equation}
A_p D_p = C v_p^2,
\label{eq:ApDp}
\end{equation}
where $C=0.64$ is a fitting parameter. If all the initial kinetic energy is transfered or dissipated by the penetration to the depth $D_p$, $C$ should be $1/2$ due to the energy balance $m_p v_p^2/2= m_p A_p D_p$, where $m_p$ is the mass of the projectile. The obtained value $C=0.64$ is close to $1/2$. However, in some cases, the final penetration depth is shallower than $D_p$. In addition, a small, but finite, projectile velocity sometimes remains after the target break-up, as shown later~(Fig.~\ref{fig:after-vel}). The scaling in Fig.~\ref{fig:ApDp-vp} includes all the impact data. Therefore, we can only discuss the global trend of the kinematics of projectile from this simple scaling analysis.

To directly estimate the degree of dissipation, we also analyzed the late stage of the impacts. In the late stage, the projectile becomes visible in some cases (see Fig.~\ref{fig:raw}(e,j)). In such cases, we can manually measure the velocity of the projectile after the break-through of the target, $v'$. The measured ratio $(v'/v_p)^2$ as a function of $v_p$ is shown in Fig.~\ref{fig:after-vel}. Since $(v'/v_p)^2$ indicates the fraction of preserved kinetic-energy after the impact, it is analogous to the squared restitution coefficient $\epsilon^2$. The velocity dependence of $\epsilon$ has been studied for viscoelastic ($\epsilon^2 \propto v^{-2/3}$)~\cite{Schwager:2008} and plastic ($\epsilon^2 \propto v^{-1/2}$)~\cite{Johnson:1985} cases. For comparisons, these relations are shown in the inset of Fig.~\ref{fig:after-vel}. The experimental data distribute below these two lines. Thus, the impact dynamics could be more dissipative than viscoelastic and plastic impacts, in terms of velocity dependence. Since the observation duration is limited, the fragmentation threshold of $v_p$ in the low-speed regime cannot be precisely estimated from this analysis. Practically, in the high-speed regime ($v_p \stackrel{>}{\sim} 0.4$~m~s$^{-1}$), the projectile loses more than 80\% 
of the initial kinetic energy by breaking through the target.

\begin{figure}
\begin{center}
\resizebox{0.35\textwidth}{!}{\includegraphics{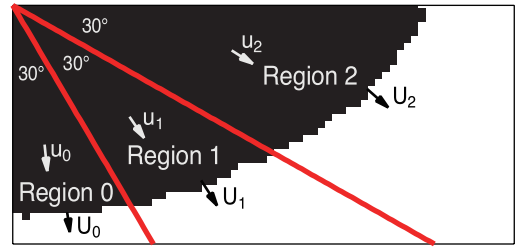}}
\end{center}
\caption{Division of the analyzed target into regions. The black part corresponds to the projected target area. From the center of the longest horizontal axis of the target image (top-left corner of the image), the target is divided into three regions by the azimuthal angle from the vertical direction. While only the right-hand side of the target is shown here, both sides are used in the actual analysis.}
\label{fig:regions}
\end{figure}

Next, we would like to focus on the target behavior. In Fig.~\ref{fig:raw}, an expansion of the target due to the impact is observed. To describe the target expansion, we divided the target into three regions as shown in Fig.~\ref{fig:regions} and analyzed the region below the horizontally longest axis of the projected target image. The regions~$0$, $1$, and $2$ correspond to the ranges of $|\theta|\leq 30^{\circ}$, $30^{\circ} < |\theta| \leq 60^{\circ}$, and $60^{\circ} < |\theta| \leq 90^{\circ}$, respectively, where $\theta$ is the azimuthal angle from the vertical direction. By image analysis (see appendix for details), the expansion velocity at the outer boundary of the target, $U_i$, and that at the center of the target profile, $u_i$, were measured for each region $i$.

Possible physical mechanisms governing the expansion rates $U_i$ and $u_i$ could be momentum or energy transfer from the projectile to the target. We tried several scaling ideas and finally found that the energy-based scaling can best explain the data behavior. Specifically, the following relations are considered,
\begin{subequations}
\label{eq:energy-scaling-all} % notice location
\begin{eqnarray}
m_iU_i^2 &=& K_i^2 \mu v_p^2,\label{eq:energy-scaling1} \\
m_iu_i^2 &=& k_i^2 \mu v_p^2,\label{eq:energy-scaling2}
\end{eqnarray}
\end{subequations}
where $K_i$ and $k_i$ are fitting parameters characterizing the energy-transfer efficiency and $\mu = m_p m_t/(m_p+m_t)$ and $m_i$ are the reduced mass and the mass of the $i$th region of the target, respectively. Here, $m_p$ and $m_t$ are the projectile and target mass, respectively. To estimate $m_i$, we assume an axisymmetric shape and a  uniform density within the target. As we can only observe the projected two-dimensional motion, we assume an axisymmetric expansion in three-dimensional space to compute the relations given by Eq.~(\ref{eq:energy-scaling-all}).

\begin{figure*}
\begin{center}
\includegraphics[width=6.5 in]{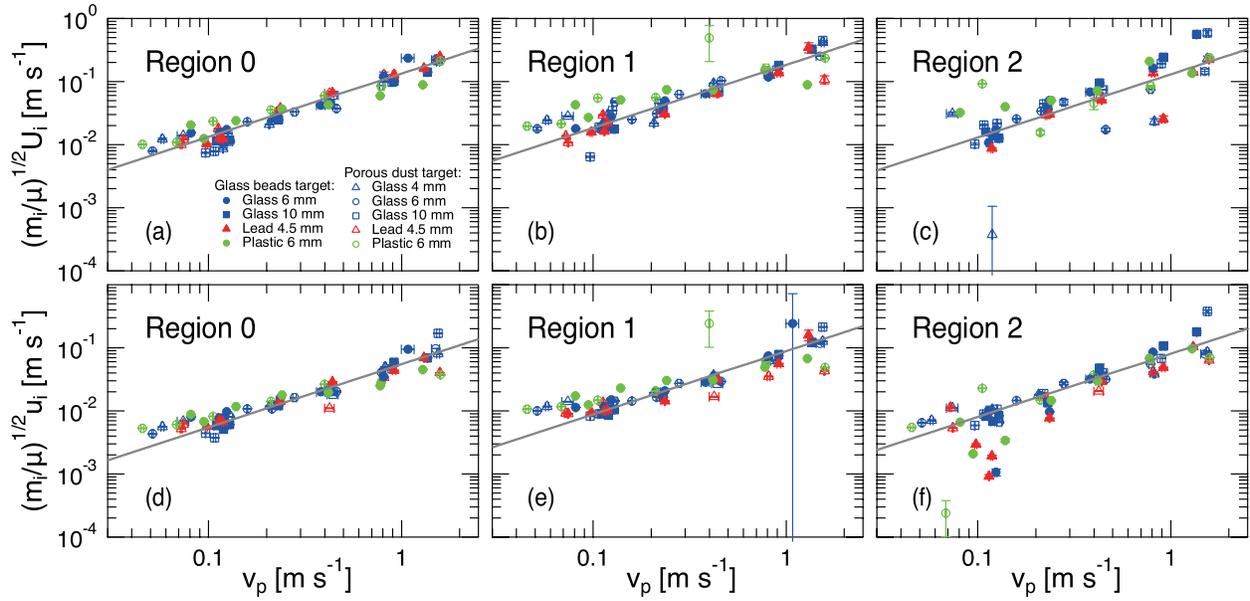}
\end{center}
\caption{Energy-based scaling of the expansion velocities at the rim, $U_i$, and the center, $u_i$, respectively. All impact data are plotted in all the panels. Open and filled symbols correspond to porous-dust-agglomerates and dense-glass-beads targets, respectively. By normalizing $U_i$ and $u_i$ using the mass ratio $(m_i/\mu)^{1/2}$, all expansion rates can be collapsed on a unified scaling relation shown by the gray lines, which are linear fittings with slope unity in the log-log plots.
}
\label{fig:energy-scaling}
\end{figure*}

The data analysis based on this energy scaling (Eq.~(\ref{eq:energy-scaling-all})) is shown in Fig.~\ref{fig:energy-scaling}. In Fig.~\ref{fig:energy-scaling}, both, the porous-dust-agglomerates and the dense-glass-beads data are collapsed on an identical unified scaling. This implies that the difference between porous-dust-agglomerates and dense-glass-beads targets can be expressed only by their respective mass (or density) difference. The deformation and fracturing of the porous-dust agglomerates within the target are negligible under the current experimental conditions. Both types of clusters consisting of soft or hard particles obey the same energy-transfer scaling. However, the data at small $v_p$ slightly deviate from the scaling. Besides this, the plastic projectile data might also show a slightly different scaling trend. These deviations could originate from the $v_p$ and projectile-type dependencies of $(v'/v_p)^2$ (Fig.~\ref{fig:after-vel}). Indeed, a systematic variation of the slopes in the inset of Fig.~\ref{fig:after-vel} can be confirmed depending on the projectile density. Nevertheless, the entire data globally obey the unified scaling relation in Fig.~\ref{fig:energy-scaling}.

The gray lines in Fig.~\ref{fig:energy-scaling} represent the least-square fittings to Eq.~(\ref{eq:energy-scaling-all}). We obtained $K_0=0.14$, $K_1=0.18$, $K_2=0.13$, $k_0=0.055$, $k_1=0.088$, and $k_2=0.081$, respectively. Although the values of $K_i$ and $k_i$ fluctuate, significant anisotropy cannot be confirmed, i.e., $K_0 \simeq K_1 \simeq K_2$ and $k_0 \simeq k_1 \simeq k_2$. This means that the impacted cluster isotropically expands from the impact point. Namely, the impact can be approximated by a point source, like an explosion. 

The total energy-transfer efficiency can be estimated by the sum of all three regions, $K^2=\sum_iK_i^2 \simeq 0.069$ and $k^2=\sum_ik_i^2 \simeq 0.017$. Since the value of $K^2$ is computed based on the free boundary (rim) motion, it represents the upper limit of energy0transfer efficiency. Therefore, the estimated energy-transfer efficiency in the early stage of expansion is approximately $2$ - $7$\%. 
We conclude that in our experiments the energy-transfer efficiency is independent of the target material and the impact velocity.

The internal total cohesion energy of the clusters can be determined by the known collision properties of the dust agglomerates and glass beads. Both particle types possess sticking threshold velocities of $v_{\rm thr}=1$~cm~s$^{-1}$ mainly because they are made of similar materials~(see \cite{Weidling:2012} and extrapolation from the data in \cite{Heisselmann:2010}). Thus, the total binding energy is $E_{\rm bind}=(1/2) Z m_t v_{\rm thr}^2$, with $Z$ being the coordination number inside the cluster. The latter can be estimated by $Z \simeq 6$. Thus, we get $E_{\rm bind}=3\times 10^{-7}$~J for the dust-agglomerate targets and $E_{\rm bind}=9\times 10^{-7}$~J for the glass-bead targets, respectively. The kinetic energy of the projectiles ranged from $E_{\rm kin}=1\times 10^{-7}$~J to $2\times 10^{-3}$~J for the dust-agglomerate clusters and $E_{\rm kin}=4\times 10^{-7}$~J to $1\times 10^{-3}$~J for the glass-bead clusters. The cohesion among the dust agglomerates and glass beads provided for a container-free experiment, because the target clusters stayed intact prior to impact, but the total binding energy is too small to play any major role in the impact dynamics.

The above-mentioned results indicate the very dissipative nature of impacts into granular media. Similar dissipative characteristics have been reported in mutual collision of macroscopic granular clusters on a two-dimensional floating setup~\cite{Burton:2013PRL,Burton:2013PRE}. According to~\cite{Burton:2013PRE}, for head-on collision, the kinetic energy is reduced to $5$ - $20$\% 
of its initial value. \cite{Burton:2013PRE} also found that the dissipative behavior is independent of the number of particles in the colliding clusters. In this study, we revealed that the dissipative nature of granular clusters is independent of the mechanical properties of the constituent particles.

To understand the energy transfer and dissipation more quantitatively, we use a model of a random-walk-like (diffusive) sequence of binary collisions~\cite{Crassous:2007,Valance:2009}. In the model, the energy-transfer efficiency per collision, $\beta^2$, is introduced, viz., the energy-transfer efficiency by $n$-time collisions is $\beta^{2n}$. Therefore, in the current system, the transfer efficiency of kinetic energy from the projectile to the outer boundary of the target can be written as 
\begin{equation}
K^2 = \beta^{2n}/\left[ 1-(v'/v_p)^2 \right].
\label{eq:collision-model}
\end{equation} 
To compute the specific value of $\beta$, we substitute $K^2=0.069$, $(v'/v_p)^2=0.15$ (average value for $v_p>0.4$~m~s$^{-1}$, see Fig.~\ref{fig:after-vel}), and $n=81=9^2$ which corresponds to the square of the average radius of the target in grain diameter units. Since the model assumes a random-walk collisional chain, the number of collisions is estimated by the square of the linear size~\cite{Valance:2009}. Then, we obtain $\beta=0.98$ which is close to the value obtained in the impact-induced splashing of hard particles~\cite{Crassous:2007,Valance:2009}. That is, the good data collapse in Fig.~\ref{fig:energy-scaling} implies that the energy-transfer efficiency per collision is virtually independent of the particle properties, impact situations, and boundary conditions. Moreover, the diffusive propagation of energy is consistent with the isotropic expansion as well.    

The current experimental results suggest that the impact-induced energy transfer and dissipation do not depend on the details of the target material. This fact is useful to build a simple model of energy partition in impact-driven porous-dust growth that is a key element of the planet formation process. However, the experimentally accessed $v_p$ range is still limited. Both, higher- and lower-speed regimes must be examined. To this end, the use of taller drop towers is a possible attempt for future research. Furthermore, the effects of composing particle shapes and their size distribution should be investigated to check the universality of the scaling proposed in this study. In the study reported here, we provide the first results on this issue by using a unique experimental setup.

In summary, we performed microgravity impact experiments using various solid projectiles and granular clusters. A vacuum drop-tower setup was used to focus on the intrinsic granular behavior by removing the effects of gravity and interstitial air. To study the impact of the constituent-particle properties, we used porous dust agglomerates and dense glass beads as particles constructing the granular targets. The expansion rate of the impacted granular cluster was measured and scaled by the energy-transfer law. Based on the analyzed results, it turned out that energy-transfer law is independent of the projectile and target properties and the expansion of the impacted cluster is isotropic. The total transferred kinetic energy from the projectile to the target is about $2$ - $7$\%. 
These behaviors can be understood by the model of dissipative diffusion of binary collisions.

\begin{acknowledgments}
We thank A. Landeck for the assistance in experimental preparation. HK thanks JSPS KAKENHI Grant Nos.~15KK0158 and 18H03679 for financial support. JB thanks the Deutsche Forschungsgemeinschaft (DFG, grant Bl 298/24-1) and the Deutsches Zentrum f\"ur Luft- und Raumfahrt (DLR, grant 50WM1536) for financial support.
\end{acknowledgments}

\bibliography{DBI}

\begin{thebibliography}{28}%
\makeatletter
\providecommand \@ifxundefined [1]{%
 \@ifx{#1\undefined}
}%
\providecommand \@ifnum [1]{%
 \ifnum #1\expandafter \@firstoftwo
 \else \expandafter \@secondoftwo
 \fi
}%
\providecommand \@ifx [1]{%
 \ifx #1\expandafter \@firstoftwo
 \else \expandafter \@secondoftwo
 \fi
}%
\providecommand \natexlab [1]{#1}%
\providecommand \enquote  [1]{``#1''}%
\providecommand \bibnamefont  [1]{#1}%
\providecommand \bibfnamefont [1]{#1}%
\providecommand \citenamefont [1]{#1}%
\providecommand \href@noop [0]{\@secondoftwo}%
\providecommand \href [0]{\begingroup \@sanitize@url \@href}%
\providecommand \@href[1]{\@@startlink{#1}\@@href}%
\providecommand \@@href[1]{\endgroup#1\@@endlink}%
\providecommand \@sanitize@url [0]{\catcode `\\12\catcode `\$12\catcode
  `\&12\catcode `\#12\catcode `\^12\catcode `\_12\catcode `\%12\relax}%
\providecommand \@@startlink[1]{}%
\providecommand \@@endlink[0]{}%
\providecommand \url  [0]{\begingroup\@sanitize@url \@url }%
\providecommand \@url [1]{\endgroup\@href {#1}{\urlprefix }}%
\providecommand \urlprefix  [0]{URL }%
\providecommand \Eprint [0]{\href }%
\providecommand \doibase [0]{http://dx.doi.org/}%
\providecommand \selectlanguage [0]{\@gobble}%
\providecommand \bibinfo  [0]{\@secondoftwo}%
\providecommand \bibfield  [0]{\@secondoftwo}%
\providecommand \translation [1]{[#1]}%
\providecommand \BibitemOpen [0]{}%
\providecommand \bibitemStop [0]{}%
\providecommand \bibitemNoStop [0]{.\EOS\space}%
\providecommand \EOS [0]{\spacefactor3000\relax}%
\providecommand \BibitemShut  [1]{\csname bibitem#1\endcsname}%
\let\auto@bib@innerbib\@empty
%</preamble>
\bibitem [{\citenamefont {Andreotti}\ \emph {et~al.}(2013)\citenamefont
  {Andreotti}, \citenamefont {Forterre},\ and\ \citenamefont
  {Pouliquen}}]{Andreotti:2013}%
  \BibitemOpen
  \bibfield  {author} {\bibinfo {author} {\bibfnamefont {B.}~\bibnamefont
  {Andreotti}}, \bibinfo {author} {\bibfnamefont {Y.}~\bibnamefont {Forterre}},
  \ and\ \bibinfo {author} {\bibfnamefont {O.}~\bibnamefont {Pouliquen}},\
  }\href@noop {} {\emph {\bibinfo {title} {{Granular Media: Between Fluid and
  Solid}}}}\ (\bibinfo  {publisher} {Cambridge University Press},\ \bibinfo
  {year} {2013})\BibitemShut {NoStop}%
\bibitem [{\citenamefont {Ruiz-Su{\'a}rez}(2013)}]{RuizSuarez:2013}%
  \BibitemOpen
  \bibfield  {author} {\bibinfo {author} {\bibfnamefont {J.~C.}\ \bibnamefont
  {Ruiz-Su{\'a}rez}},\ }\href@noop {} {\bibfield  {journal} {\bibinfo
  {journal} {Rep. Prog. Phys.}\ }\textbf {\bibinfo {volume} {76}},\ \bibinfo
  {pages} {066601} (\bibinfo {year} {2013})}\BibitemShut {NoStop}%
\bibitem [{\citenamefont {Katsuragi}(2016)}]{Katsuragi:2016}%
  \BibitemOpen
  \bibfield  {author} {\bibinfo {author} {\bibfnamefont {H.}~\bibnamefont
  {Katsuragi}},\ }\href@noop {} {\emph {\bibinfo {title} {{Physics of Soft
  Impact and Cratering}}}}\ (\bibinfo  {publisher} {Springer},\ \bibinfo {year}
  {2016})\BibitemShut {NoStop}%
\bibitem [{\citenamefont {van~der Meer}(2017)}]{vanderMeer:2017}%
  \BibitemOpen
  \bibfield  {author} {\bibinfo {author} {\bibfnamefont {D.}~\bibnamefont
  {van~der Meer}},\ }\href@noop {} {\bibfield  {journal} {\bibinfo  {journal}
  {Annu. Rev. Fluid Mech.}\ }\textbf {\bibinfo {volume} {49}},\ \bibinfo
  {pages} {463} (\bibinfo {year} {2017})}\BibitemShut {NoStop}%
\bibitem [{\citenamefont {Katsuragi}\ and\ \citenamefont
  {Durian}(2007)}]{Katsuragi:2007}%
  \BibitemOpen
  \bibfield  {author} {\bibinfo {author} {\bibfnamefont {H.}~\bibnamefont
  {Katsuragi}}\ and\ \bibinfo {author} {\bibfnamefont {D.~J.}\ \bibnamefont
  {Durian}},\ }\href@noop {} {\bibfield  {journal} {\bibinfo  {journal} {Nat.
  Phys.}\ }\textbf {\bibinfo {volume} {3}},\ \bibinfo {pages} {420} (\bibinfo
  {year} {2007})}\BibitemShut {NoStop}%
\bibitem [{\citenamefont {Katsuragi}\ and\ \citenamefont
  {Durian}(2013)}]{Katsuragi:2013}%
  \BibitemOpen
  \bibfield  {author} {\bibinfo {author} {\bibfnamefont {H.}~\bibnamefont
  {Katsuragi}}\ and\ \bibinfo {author} {\bibfnamefont {D.~J.}\ \bibnamefont
  {Durian}},\ }\href@noop {} {\bibfield  {journal} {\bibinfo  {journal} {Phys.
  Rev. E}\ }\textbf {\bibinfo {volume} {87}},\ \bibinfo {pages} {052208}
  (\bibinfo {year} {2013})}\BibitemShut {NoStop}%
\bibitem [{\citenamefont {Clark}\ \emph {et~al.}(2014)\citenamefont {Clark},
  \citenamefont {Petersen},\ and\ \citenamefont {Behringer}}]{Clark:2014}%
  \BibitemOpen
  \bibfield  {author} {\bibinfo {author} {\bibfnamefont {A.~H.}\ \bibnamefont
  {Clark}}, \bibinfo {author} {\bibfnamefont {A.~J.}\ \bibnamefont {Petersen}},
  \ and\ \bibinfo {author} {\bibfnamefont {R.~P.}\ \bibnamefont {Behringer}},\
  }\href {\doibase 10.1103/PhysRevE.89.012201} {\bibfield  {journal} {\bibinfo
  {journal} {Phys. Rev. E}\ }\textbf {\bibinfo {volume} {89}},\ \bibinfo
  {pages} {012201} (\bibinfo {year} {2014})}\BibitemShut {NoStop}%
\bibitem [{\citenamefont {Bester}\ and\ \citenamefont
  {Behringer}(2017)}]{Bester:2017}%
  \BibitemOpen
  \bibfield  {author} {\bibinfo {author} {\bibfnamefont {C.~S.}\ \bibnamefont
  {Bester}}\ and\ \bibinfo {author} {\bibfnamefont {R.~P.}\ \bibnamefont
  {Behringer}},\ }\href@noop {} {\bibfield  {journal} {\bibinfo  {journal}
  {Phys. Rev. E}\ }\textbf {\bibinfo {volume} {95}},\ \bibinfo {pages} {032906}
  (\bibinfo {year} {2017})}\BibitemShut {NoStop}%
\bibitem [{\citenamefont {Walsh}\ \emph {et~al.}(2003)\citenamefont {Walsh},
  \citenamefont {Holloway}, \citenamefont {Habdas},\ and\ \citenamefont
  {de~Bruyn}}]{Walsh:2003}%
  \BibitemOpen
  \bibfield  {author} {\bibinfo {author} {\bibfnamefont {A.~M.}\ \bibnamefont
  {Walsh}}, \bibinfo {author} {\bibfnamefont {K.~E.}\ \bibnamefont {Holloway}},
  \bibinfo {author} {\bibfnamefont {P.}~\bibnamefont {Habdas}}, \ and\ \bibinfo
  {author} {\bibfnamefont {J.~R.}\ \bibnamefont {de~Bruyn}},\ }\href@noop {}
  {\bibfield  {journal} {\bibinfo  {journal} {Phys. Rev. Lett.}\ }\textbf
  {\bibinfo {volume} {91}},\ \bibinfo {pages} {104301} (\bibinfo {year}
  {2003})}\BibitemShut {NoStop}%
\bibitem [{\citenamefont {Wada}\ \emph {et~al.}(2006)\citenamefont {Wada},
  \citenamefont {Senshu},\ and\ \citenamefont {Matsui}}]{Wada:2006}%
  \BibitemOpen
  \bibfield  {author} {\bibinfo {author} {\bibfnamefont {K.}~\bibnamefont
  {Wada}}, \bibinfo {author} {\bibfnamefont {H.}~\bibnamefont {Senshu}}, \ and\
  \bibinfo {author} {\bibfnamefont {T.}~\bibnamefont {Matsui}},\ }\href@noop {}
  {\bibfield  {journal} {\bibinfo  {journal} {Icarus}\ }\textbf {\bibinfo
  {volume} {180}},\ \bibinfo {pages} {528} (\bibinfo {year}
  {2006})}\BibitemShut {NoStop}%
\bibitem [{\citenamefont {Crassous}\ \emph {et~al.}(2007)\citenamefont
  {Crassous}, \citenamefont {Beladjine},\ and\ \citenamefont
  {Valance}}]{Crassous:2007}%
  \BibitemOpen
  \bibfield  {author} {\bibinfo {author} {\bibfnamefont {J.}~\bibnamefont
  {Crassous}}, \bibinfo {author} {\bibfnamefont {D.}~\bibnamefont {Beladjine}},
  \ and\ \bibinfo {author} {\bibfnamefont {A.}~\bibnamefont {Valance}},\
  }\href@noop {} {\bibfield  {journal} {\bibinfo  {journal} {Phys. Rev. Lett.}\
  }\textbf {\bibinfo {volume} {99}},\ \bibinfo {pages} {248001} (\bibinfo
  {year} {2007})}\BibitemShut {NoStop}%
\bibitem [{\citenamefont {Valance}\ and\ \citenamefont
  {Crassous}(2009)}]{Valance:2009}%
  \BibitemOpen
  \bibfield  {author} {\bibinfo {author} {\bibfnamefont {A.}~\bibnamefont
  {Valance}}\ and\ \bibinfo {author} {\bibfnamefont {J.}~\bibnamefont
  {Crassous}},\ }\href@noop {} {\bibfield  {journal} {\bibinfo  {journal}
  {Euro. Phys. J. E}\ }\textbf {\bibinfo {volume} {30}},\ \bibinfo {pages} {43}
  (\bibinfo {year} {2009})}\BibitemShut {NoStop}%
\bibitem [{\citenamefont {Deboeuf}\ \emph {et~al.}(2009)\citenamefont
  {Deboeuf}, \citenamefont {Gondret},\ and\ \citenamefont
  {Rabaud}}]{Deboeuf:2009}%
  \BibitemOpen
  \bibfield  {author} {\bibinfo {author} {\bibfnamefont {S.}~\bibnamefont
  {Deboeuf}}, \bibinfo {author} {\bibfnamefont {P.}~\bibnamefont {Gondret}}, \
  and\ \bibinfo {author} {\bibfnamefont {M.}~\bibnamefont {Rabaud}},\
  }\href@noop {} {\bibfield  {journal} {\bibinfo  {journal} {Phys. Rev. E}\
  }\textbf {\bibinfo {volume} {79}},\ \bibinfo {pages} {041306} (\bibinfo
  {year} {2009})}\BibitemShut {NoStop}%
\bibitem [{\citenamefont {Tanabe}\ \emph {et~al.}(2017)\citenamefont {Tanabe},
  \citenamefont {Shimada}, \citenamefont {Ito},\ and\ \citenamefont
  {Nishimori}}]{Tanabe:2017}%
  \BibitemOpen
  \bibfield  {author} {\bibinfo {author} {\bibfnamefont {T.}~\bibnamefont
  {Tanabe}}, \bibinfo {author} {\bibfnamefont {T.}~\bibnamefont {Shimada}},
  \bibinfo {author} {\bibfnamefont {N.}~\bibnamefont {Ito}}, \ and\ \bibinfo
  {author} {\bibfnamefont {H.}~\bibnamefont {Nishimori}},\ }\href@noop {}
  {\bibfield  {journal} {\bibinfo  {journal} {Phys. Rev. E}\ }\textbf {\bibinfo
  {volume} {95}},\ \bibinfo {pages} {022906} (\bibinfo {year}
  {2017})}\BibitemShut {NoStop}%
\bibitem [{\citenamefont {Royer}\ \emph {et~al.}(2009)\citenamefont {Royer},
  \citenamefont {Evans}, \citenamefont {Oyarte}, \citenamefont {Guo},
  \citenamefont {Kapit}, \citenamefont {M{\"o}bius}, \citenamefont
  {Waitukaitis},\ and\ \citenamefont {Jaeger}}]{Royer:2009}%
  \BibitemOpen
  \bibfield  {author} {\bibinfo {author} {\bibfnamefont {J.~R.}\ \bibnamefont
  {Royer}}, \bibinfo {author} {\bibfnamefont {D.~J.}\ \bibnamefont {Evans}},
  \bibinfo {author} {\bibfnamefont {L.}~\bibnamefont {Oyarte}}, \bibinfo
  {author} {\bibfnamefont {Q.}~\bibnamefont {Guo}}, \bibinfo {author}
  {\bibfnamefont {E.}~\bibnamefont {Kapit}}, \bibinfo {author} {\bibfnamefont
  {M.~E.}\ \bibnamefont {M{\"o}bius}}, \bibinfo {author} {\bibfnamefont
  {S.~R.}\ \bibnamefont {Waitukaitis}}, \ and\ \bibinfo {author} {\bibfnamefont
  {H.~M.}\ \bibnamefont {Jaeger}},\ }\href@noop {} {\bibfield  {journal}
  {\bibinfo  {journal} {Nature}\ }\textbf {\bibinfo {volume} {459}},\ \bibinfo
  {pages} {1110} (\bibinfo {year} {2009})}\BibitemShut {NoStop}%
\bibitem [{\citenamefont {Blum}(2018)}]{Blum:2018}%
  \BibitemOpen
  \bibfield  {author} {\bibinfo {author} {\bibfnamefont {J.}~\bibnamefont
  {Blum}},\ }\href@noop {} {\bibfield  {journal} {\bibinfo  {journal} {Space
  Sci. Rev.}\ }\textbf {\bibinfo {volume} {214}},\ \bibinfo {pages} {52}
  (\bibinfo {year} {2018})}\BibitemShut {NoStop}%
\bibitem [{\citenamefont {G{\"u}ttler}\ \emph {et~al.}(2009)\citenamefont
  {G{\"u}ttler}, \citenamefont {Krause}, \citenamefont {Geretshauser},
  \citenamefont {Speith},\ and\ \citenamefont {Blum}}]{Guttler:2009}%
  \BibitemOpen
  \bibfield  {author} {\bibinfo {author} {\bibfnamefont {C.}~\bibnamefont
  {G{\"u}ttler}}, \bibinfo {author} {\bibfnamefont {M.}~\bibnamefont {Krause}},
  \bibinfo {author} {\bibfnamefont {R.~J.}\ \bibnamefont {Geretshauser}},
  \bibinfo {author} {\bibfnamefont {R.}~\bibnamefont {Speith}}, \ and\ \bibinfo
  {author} {\bibfnamefont {J.}~\bibnamefont {Blum}},\ }\href@noop {} {\bibfield
   {journal} {\bibinfo  {journal} {ApJ}\ }\textbf {\bibinfo {volume} {701}},\
  \bibinfo {pages} {130} (\bibinfo {year} {2009})}\BibitemShut {NoStop}%
\bibitem [{\citenamefont {Katsuragi}\ and\ \citenamefont
  {Blum}(2017)}]{Katsuragi:2017}%
  \BibitemOpen
  \bibfield  {author} {\bibinfo {author} {\bibfnamefont {H.}~\bibnamefont
  {Katsuragi}}\ and\ \bibinfo {author} {\bibfnamefont {J.}~\bibnamefont
  {Blum}},\ }\href@noop {} {\bibfield  {journal} {\bibinfo  {journal} {ApJ}\
  }\textbf {\bibinfo {volume} {851}},\ \bibinfo {pages} {23} (\bibinfo {year}
  {2017})}\BibitemShut {NoStop}%
\bibitem [{\citenamefont {Whizin}\ \emph {et~al.}(2017)\citenamefont {Whizin},
  \citenamefont {Blum},\ and\ \citenamefont {Colwell}}]{Whizin:2017}%
  \BibitemOpen
  \bibfield  {author} {\bibinfo {author} {\bibfnamefont {A.~D.}\ \bibnamefont
  {Whizin}}, \bibinfo {author} {\bibfnamefont {J.}~\bibnamefont {Blum}}, \ and\
  \bibinfo {author} {\bibfnamefont {J.~E.}\ \bibnamefont {Colwell}},\
  }\href@noop {} {\bibfield  {journal} {\bibinfo  {journal} {ApJ}\ }\textbf
  {\bibinfo {volume} {836}},\ \bibinfo {pages} {94} (\bibinfo {year}
  {2017})}\BibitemShut {NoStop}%
\bibitem [{\citenamefont {Murdoch}\ \emph {et~al.}(2017)\citenamefont
  {Murdoch}, \citenamefont {Avila~Martinez}, \citenamefont {Sunday},
  \citenamefont {Zenou}, \citenamefont {Cherrier}, \citenamefont {Cadu},\ and\
  \citenamefont {Gourinat}}]{Murdoch:2017}%
  \BibitemOpen
  \bibfield  {author} {\bibinfo {author} {\bibfnamefont {N.}~\bibnamefont
  {Murdoch}}, \bibinfo {author} {\bibfnamefont {I.}~\bibnamefont
  {Avila~Martinez}}, \bibinfo {author} {\bibfnamefont {C.}~\bibnamefont
  {Sunday}}, \bibinfo {author} {\bibfnamefont {E.}~\bibnamefont {Zenou}},
  \bibinfo {author} {\bibfnamefont {O.}~\bibnamefont {Cherrier}}, \bibinfo
  {author} {\bibfnamefont {A.}~\bibnamefont {Cadu}}, \ and\ \bibinfo {author}
  {\bibfnamefont {Y.}~\bibnamefont {Gourinat}},\ }\href@noop {} {\bibfield
  {journal} {\bibinfo  {journal} {MNRAS}\ }\textbf {\bibinfo {volume} {468}},\
  \bibinfo {pages} {1259} (\bibinfo {year} {2017})}\BibitemShut {NoStop}%
\bibitem [{\citenamefont {Planes}\ \emph {et~al.}(2017)\citenamefont {Planes},
  \citenamefont {Mill{\'a}n}, \citenamefont {Urbassek},\ and\ \citenamefont
  {Bringa}}]{Planes:2017}%
  \BibitemOpen
  \bibfield  {author} {\bibinfo {author} {\bibfnamefont {M.~B.}\ \bibnamefont
  {Planes}}, \bibinfo {author} {\bibfnamefont {E.~N.}\ \bibnamefont
  {Mill{\'a}n}}, \bibinfo {author} {\bibfnamefont {H.~M.}\ \bibnamefont
  {Urbassek}}, \ and\ \bibinfo {author} {\bibfnamefont {E.~M.}\ \bibnamefont
  {Bringa}},\ }\href@noop {} {\bibfield  {journal} {\bibinfo  {journal}
  {A{\&}A}\ }\textbf {\bibinfo {volume} {607}},\ \bibinfo {pages} {A19}
  (\bibinfo {year} {2017})}\BibitemShut {NoStop}%
\bibitem [{\citenamefont {Blum}\ \emph {et~al.}(2014)\citenamefont {Blum},
  \citenamefont {Beitz}, \citenamefont {Bukhari}, \citenamefont {Gundlach},
  \citenamefont {Hagemann}, \citenamefont {Hei{\ss}elmann}, \citenamefont
  {Kothe}, \citenamefont {Schr{\"a}pler}, \citenamefont {von Borstel},\ and\
  \citenamefont {Weidling}}]{Blum:2014}%
  \BibitemOpen
  \bibfield  {author} {\bibinfo {author} {\bibfnamefont {J.}~\bibnamefont
  {Blum}}, \bibinfo {author} {\bibfnamefont {E.}~\bibnamefont {Beitz}},
  \bibinfo {author} {\bibfnamefont {M.}~\bibnamefont {Bukhari}}, \bibinfo
  {author} {\bibfnamefont {B.}~\bibnamefont {Gundlach}}, \bibinfo {author}
  {\bibfnamefont {J.-H.}\ \bibnamefont {Hagemann}}, \bibinfo {author}
  {\bibfnamefont {D.}~\bibnamefont {Hei{\ss}elmann}}, \bibinfo {author}
  {\bibfnamefont {S.}~\bibnamefont {Kothe}}, \bibinfo {author} {\bibfnamefont
  {R.}~\bibnamefont {Schr{\"a}pler}}, \bibinfo {author} {\bibfnamefont
  {I.}~\bibnamefont {von Borstel}}, \ and\ \bibinfo {author} {\bibfnamefont
  {R.}~\bibnamefont {Weidling}},\ }\href@noop {} {\bibfield  {journal}
  {\bibinfo  {journal} {JoVE (Journal of Visualized Experiments)}\ ,\ \bibinfo
  {pages} {e51541}} (\bibinfo {year} {2014})}\BibitemShut {NoStop}%
\bibitem [{\citenamefont {Schwager}\ and\ \citenamefont
  {P{\"o}schel}(2008)}]{Schwager:2008}%
  \BibitemOpen
  \bibfield  {author} {\bibinfo {author} {\bibfnamefont {T.}~\bibnamefont
  {Schwager}}\ and\ \bibinfo {author} {\bibfnamefont {T.}~\bibnamefont
  {P{\"o}schel}},\ }\href@noop {} {\bibfield  {journal} {\bibinfo  {journal}
  {Phys. Rev. E}\ }\textbf {\bibinfo {volume} {78}},\ \bibinfo {pages} {051304}
  (\bibinfo {year} {2008})}\BibitemShut {NoStop}%
\bibitem [{\citenamefont {Johnson}(1985)}]{Johnson:1985}%
  \BibitemOpen
  \bibfield  {author} {\bibinfo {author} {\bibfnamefont {K.~L.}\ \bibnamefont
  {Johnson}},\ }\href@noop {} {\emph {\bibinfo {title} {{Contact Mechanics}}}}\
  (\bibinfo  {publisher} {Cambridge Univ. Press},\ \bibinfo {year}
  {1985})\BibitemShut {NoStop}%
\bibitem [{\citenamefont {Weidling}\ \emph {et~al.}(2012)\citenamefont
  {Weidling}, \citenamefont {G{\"u}ttler},\ and\ \citenamefont
  {Blum}}]{Weidling:2012}%
  \BibitemOpen
  \bibfield  {author} {\bibinfo {author} {\bibfnamefont {R.}~\bibnamefont
  {Weidling}}, \bibinfo {author} {\bibfnamefont {C.}~\bibnamefont
  {G{\"u}ttler}}, \ and\ \bibinfo {author} {\bibfnamefont {J.}~\bibnamefont
  {Blum}},\ }\href@noop {} {\bibfield  {journal} {\bibinfo  {journal} {Icarus}\
  }\textbf {\bibinfo {volume} {218}},\ \bibinfo {pages} {688} (\bibinfo {year}
  {2012})}\BibitemShut {NoStop}%
\bibitem [{\citenamefont {Hei{\ss}elmann}\ \emph {et~al.}(2010)\citenamefont
  {Hei{\ss}elmann}, \citenamefont {Blum}, \citenamefont {Fraser},\ and\
  \citenamefont {Wolling}}]{Heisselmann:2010}%
  \BibitemOpen
  \bibfield  {author} {\bibinfo {author} {\bibfnamefont {D.}~\bibnamefont
  {Hei{\ss}elmann}}, \bibinfo {author} {\bibfnamefont {J.}~\bibnamefont
  {Blum}}, \bibinfo {author} {\bibfnamefont {H.~J.}\ \bibnamefont {Fraser}}, \
  and\ \bibinfo {author} {\bibfnamefont {K.}~\bibnamefont {Wolling}},\
  }\href@noop {} {\bibfield  {journal} {\bibinfo  {journal} {Icarus}\ }\textbf
  {\bibinfo {volume} {206}},\ \bibinfo {pages} {424} (\bibinfo {year}
  {2010})}\BibitemShut {NoStop}%
\bibitem [{\citenamefont {Burton}\ \emph
  {et~al.}(2013{\natexlab{a}})\citenamefont {Burton}, \citenamefont {Lu},\ and\
  \citenamefont {Nagel}}]{Burton:2013PRL}%
  \BibitemOpen
  \bibfield  {author} {\bibinfo {author} {\bibfnamefont {J.~C.}\ \bibnamefont
  {Burton}}, \bibinfo {author} {\bibfnamefont {P.~Y.}\ \bibnamefont {Lu}}, \
  and\ \bibinfo {author} {\bibfnamefont {S.~R.}\ \bibnamefont {Nagel}},\
  }\href@noop {} {\bibfield  {journal} {\bibinfo  {journal} {Phys. Rev. Lett.}\
  }\textbf {\bibinfo {volume} {111}},\ \bibinfo {pages} {188001} (\bibinfo
  {year} {2013}{\natexlab{a}})}\BibitemShut {NoStop}%
\bibitem [{\citenamefont {Burton}\ \emph
  {et~al.}(2013{\natexlab{b}})\citenamefont {Burton}, \citenamefont {Lu},\ and\
  \citenamefont {Nagel}}]{Burton:2013PRE}%
  \BibitemOpen
  \bibfield  {author} {\bibinfo {author} {\bibfnamefont {J.~C.}\ \bibnamefont
  {Burton}}, \bibinfo {author} {\bibfnamefont {P.~Y.}\ \bibnamefont {Lu}}, \
  and\ \bibinfo {author} {\bibfnamefont {S.~R.}\ \bibnamefont {Nagel}},\
  }\href@noop {} {\bibfield  {journal} {\bibinfo  {journal} {Phys. Rev. E}\
  }\textbf {\bibinfo {volume} {88}},\ \bibinfo {pages} {062204} (\bibinfo
  {year} {2013}{\natexlab{b}})}\BibitemShut {NoStop}%
\end{thebibliography}%

\clearpage

\section*{Appendix (Supplemental Material)}
\subsection*{Kinematics of projectile motion}
\begin{figure}
\begin{center}
\includegraphics{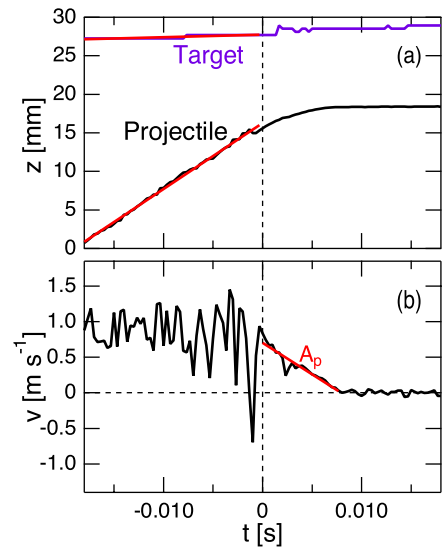}
\end{center}
\caption{Kinematic data for the impact of a $D_p=4$~mm glass projectile into a porous-dust-agglomerate cluster. The corresponding raw images are shown in Fig.~\ref{fig:raw}(a-e). he data after $t \simeq 8$~ms are not meaningful because the projectile cannot be identified thereafter.}
\label{fig:dust-kinematic}
\end{figure}

From the raw images (as, e.g., shown in Fig.~\ref{fig:raw}), the motion of projectile and target can be computed by image analysis. First, the vertical position (upper edge) of the projectile was simply detected by using a brightness threshold. Then, the vertical position of the target was defined by the level at which the horizontal diameter of the target shows a maximum. The measured positions of projectile and target for the case of Fig.~\ref{fig:raw}(a-e) are shown in Fig.~\ref{fig:dust-kinematic}(a). To reduce the noise in the data, we computed the velocities of projectile and target relative to the (free falling) camera by a linear fitting as shown in Fig.~\ref{fig:dust-kinematic}(a). From the difference of these velocities, the relative impact velocity, $v_p$, was estimated. The impact moment was directly identified by inspection of the raw images. As can be seen in Fig.~\ref{fig:dust-kinematic}(a), the velocity of the target is very small since target and camera were released to free fall simultaneously. The instantaneous relative velocity between projectile and target after the collision was computed directly from the respective position data as shown in Fig.~\ref{fig:dust-kinematic}(b). Although the data quality is not very good, we can confirm the impact-induced deceleration. However, it is difficult to estimate the time-resolved deceleration dynamics from such noisy data. Much more precise measurements, both in terms of temporal and spatial resolution, are necessary to discuss the time-resolved impact dynamics. Thus, in this study, the average deceleration $A_p$ was simply estimated by linear fitting of the decreasing relative velocity, as shown in Fig.~\ref{fig:dust-kinematic}(b).

\begin{figure}
\begin{center}
\includegraphics{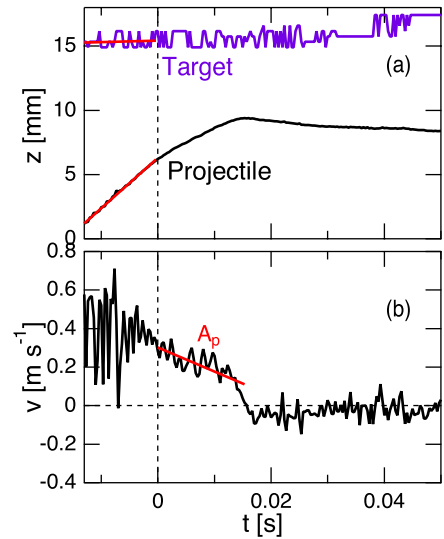}
\end{center}
\caption{Kinematic data for the impact of a $D_p=6$~mm glass projectile into a dense-glass-bead cluster. The corresponding raw images are shown in Fig.~\ref{fig:raw}(f-j). he data after $t \simeq 16$~ms are not meaningful because the projectile cannot be identified thereafter.}
\label{fig:beads-kinematic}
\end{figure}

The same analysis method was applied to all other data sets. As another example, we show here the analysis of the collision shown in Fig.~\ref{fig:raw}(f-j) (see Fig.~\ref{fig:beads-kinematic}). The qualitative behavior shown in Fig.~\ref{fig:beads-kinematic} is similar to Fig.~\ref{fig:dust-kinematic}. A minor difference that can be observed is the pseudo sudden stop of the projectile, which results in a step-like velocity jump. $v=0$ in Figs.~\ref{fig:dust-kinematic} and \ref{fig:beads-kinematic} actually corresponds to the invisibility of the projectile rather than the actual stopping of the projectile. It is impossible to follow the projectile motion after the complete penetration. Therefore, we observe a jump of $v(t)$ in some data. 
All other impact data were also analyzed by an identical image analysis to derive $v_p$ and $A_p$.

\subsection*{Expansion rate estimate}
As shown in Fig.~\ref{fig:regions}, the projected target image is divided into three regions by the azimuthal angle $\theta$ from the vertical direction; region~0:~$|\theta|\leq 30^{\circ}$, region~1:~$30^{\circ} < |\theta| \leq 60^{\circ}$, and region~2:~$60^{\circ} < |\theta| \leq 90^{\circ}$. For every time step of the high-speed images, the density of the projected target area was measured as a function of the radial distance from the origin of the cluster coordinate (top-left corner in Fig.~\ref{fig:regions}). Here, the density means the average probability to find particles (black regions in Fig.~\ref{fig:regions}) at a given radial distance.

\begin{figure}
\begin{center}
\includegraphics{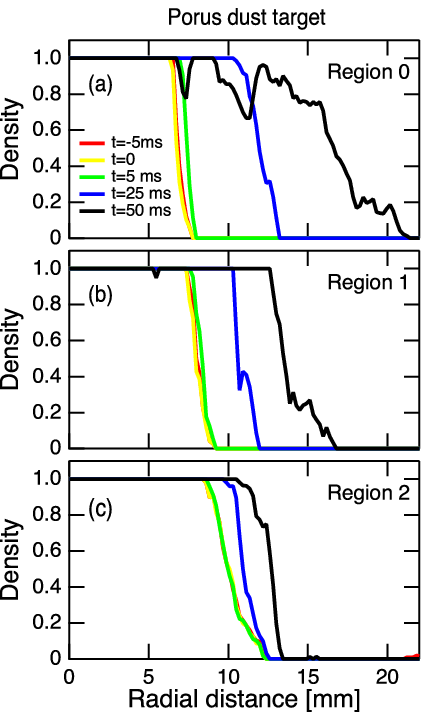}
\includegraphics{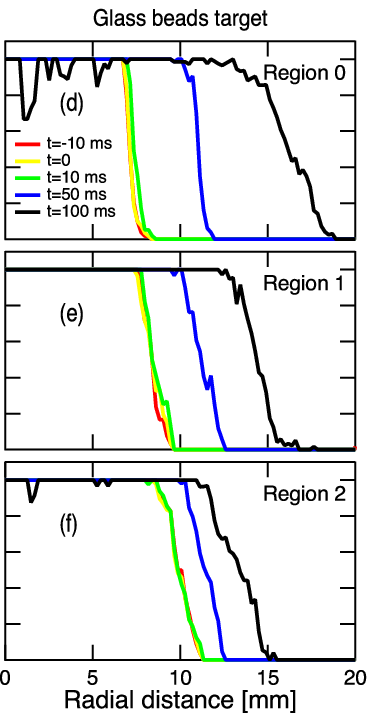}
\end{center}
\caption{Density profiles of the impacted porous-dust-agglomerates target ((a) region 0, (b) region 1, and (c) region 2) and dense-glass-beads target ((d) region 0, (e) region 1, and (f) region 2). The corresponding raw data are shown in Fig.~1(a-e) for porous dust and Fig.~1(f-j) for glass beads. The color distinguishes the different temporal snapshots.}
\label{fig:density}
\end{figure}

The measured density profiles for the three regions are shown in Fig.~\ref{fig:density}. The left and right columns correspond to porous-dust-agglomerate and dense-glass-beads targets, respectively. Obviously, the density profiles does not vary prior to the impact ($t \leq 0$). At this stage, the density profile should ideally be a step function if the target shape is spherical. In this sense, a spherical shape is ideal for the analysis of the target expansion. However, the actual shape is not spherical so that the density profiles have slopes at the target boundary, although this effect is not very significant. After the impact, the profiles move outward until finally the fragmentation (dips in the density profile) can be confirmed at the late stage of impact. To characterize the target expansion dynamics, the temporal evolution of the density profiles were analyzed. Here, we defined the cluster rim $R_i(t)$ by the position at which the density level exceeds $0.1$. The center of the density profile $r_i(t)$ was defined by the centroid of the density profile, where $i=0$, $1$, and $2$ indicates the region number.

\begin{figure}[t]
\vspace{0.5cm}
\begin{center}
\includegraphics{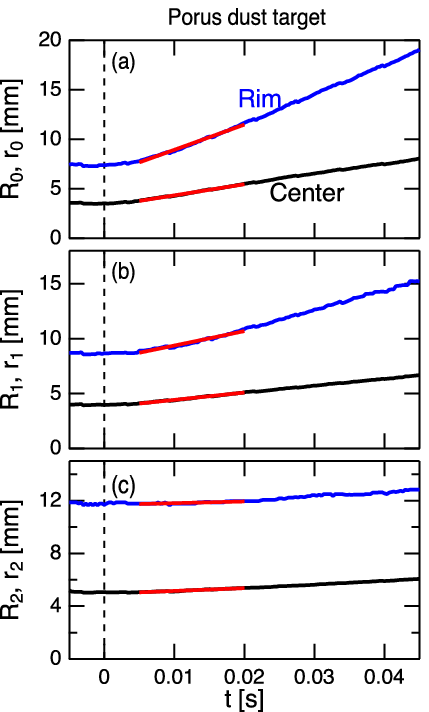}
\includegraphics{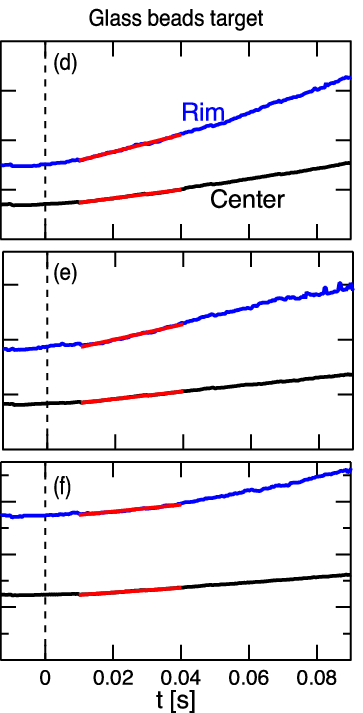}
\end{center}
\caption{Expansion of the rim position, $R_i$, and the center position, $r_i$, of the impacted porous-dust-agglomerates (a-c) and dense-glass-beads target (d-f). Data shown in Fig.~\ref{fig:density} were used to compute $R_i$ and $r_i$. The expansion rates, $U_i$ and $u_i$, were computed by linear fitting, as shown by the red lines.}
\label{fig:rx_Rx}
\end{figure}

The derived values for the target rim and center, $R_i(t)$ and $r_i(t)$, are shown in Fig.~\ref{fig:rx_Rx}. The almost linear expansion of both, rim and center, can be confirmed. We measured the expansion rate by least-squares fitting of the linearly increasing regime. While the actual fitting duration depends on the experimental conditions, the expansion rate is not very sensitive to the fitting range, as seen in Fig.~\ref{fig:rx_Rx}. In the late stage of impact (after the fragmentation of target), the behavior of $R_i(t)$ and $r_i(t)$ becomes unstable. In addition, the analysis of the late-stage behavior for small impact velocities $v_p$ is impossible, due to instrumental limitations. Therefore, we focused only on the early expansion rate in the current expansion-rate analysis. The fitted expansion rates are denoted as $U_i=dR_i/dt$ and $u_i=dr_i/dt$. The measured $U_i$ and $u_i$ data are scaled and plotted in Fig.~\ref{fig:energy-scaling}.

\end{document}